\documentclass[aps,10pt,twocolumn,superscriptaddress]{revtex4-1}
\usepackage{hyperref}
\usepackage{enumerate}
\usepackage{amsmath}
\usepackage{graphicx}
\usepackage{graphics}
\usepackage{dcolumn}%
\usepackage{bm}%
\usepackage{color}
\usepackage{subfigure}
\usepackage{float}
\usepackage{docs}%
 \usepackage[]{lineno} 
\begin{document}

\title{Free Expansion of Yukawa Gas in the Constant Plasma Background} 

\author{Manish K. Shukla} \email[Email: shuklamanish786@gmail.com]\
\affiliation{Department of Physics and Astrophysics, University of Delhi, Delhi 110007, India}
\affiliation{Department of Physics, Dyal Singh College, Delhi 110003, India}

\author{K. Avinash}
\email[Email: khareavinash82@gmail.com] \
\affiliation{Sikkim University, 6th Mile, Gangtok, 737102, India}


\begin{abstract}  
We discuss the irreversible free expansion phenomenon for Yukawa gas and obtain the corresponding change in the temperature of gas. In our expansion scheme, the system is not allowed to exchange heat with surroundings during free expansion, therefore, present expansion scheme refers to \emph{adiabatic} free expansion. Using first principle classical Molecular Dynamics (MD) simulation with reflecting boundary conditions, we show that the Yukawa gases exhibits heating effect during the process of adiabatic free expansion. We  also obtain the scaling for change in temperature and establish that the change in temperature is directly proportional to the change in number density of gas. The scaling for temperature is also obtained analytically taking the mean field limit of thermodynamic model proposed by Avinash. The simulation results are consistent with the analytical results. 

\end{abstract}

\keywords{ Dusty plasma, Molecular dynamics, Free expansion, Thermodynamics }


\maketitle 

\section{Introduction\label{Sec. I}}

Free expansion (also known as Joule expansion) is a non quasi-static irreversible phenomena where gas is allowed to expand in vacuum (for vacuum, external pressure $P_{ext}=0$) from one equilibrium state to other equilibrium state through a series of non equilibrium intermediate states. In the free expansion process the system is thermally isolated and heat is neither allowed to enter nor leave the system (i.e. $\Delta Q=0$). Further, in the process of free expansion the system is also isolated mechanically, hence, the net work done by the gas is zero, therefore, the total internal energy of the gas remains unchanged i.e., $\Delta U=0$. The free expansion of ideal and real gases is an interesting thermodynamic phenomenon. For ideal gases there is no interaction among particles, the internal energy $U$ is function of temperature $T$ only, therefore, in the free expansion process $\Delta T=0$, i.e., temperature remains constant. However, for real gases since the interaction between the molecules is a function of their relative distances, the internal energy is not only the function of temperature but also the function of volume, i.e. $U=U(T,V)$. For real gases the change in temperature is calculated using Joule's coefficient $J$ defined as, $ J=(\partial T/\partial V)_U$.

Dusty/Complex plasma is a four component plasma where micron/submicron sized particles are dispersed in usual electron-ion plasm along with neutrals gas atoms. The typical mass of the dust grains ranges from $10^{10}-10^{12} $ times the mass of proton. 
In nature, a variety of astrophysical systems such as interstellar clouds, solar system, planetary rings, cometary tails, Earth's environment etc. contain micron to sub-micron sized dust particles in different amounts \cite{frank2000waves}. Dusty plasma can also be formed in laboratories for example in DC and RF discharges. 
In the presence of gaseous plasma, these macroscopic dust particles become negatively charged and start interacting with each other, as well as with ions and electrons via long-range electric fields giving rise to collective behavior \cite{shukla2009colloquium}. This leads to the formation of special medium called the ``dusty plasma''. 
For a typical micron sized dust grain, the charge could be of the order $10^{3}-10^{5} $ times the electronic charge. 
The presence of electron ion plasma in the background of negatively charged dust grains also screens the electrostatic potential acting among dust grains. Consequently, the electrostatic interaction for dust grains is governed rather by Yukawa (or screened Coulomb) potential  which has the form,
\begin{equation}
\phi(r) = \frac{Q_d}{4\pi\epsilon_0} \frac{\exp(-r/\lambda_D)}{r} 
\label{Eq:1}
\end{equation}
where $Q_d$ is dust charge, $r$ is the distance between two dust particles and $\lambda_D$ is screening length.
The screening length depends on density and temperature of the background plasma as $ 1/\lambda_D^2=\left(\frac{e^2n_{e0}}{\epsilon_0k_BT_e}+ \frac{q^2 n_{i0}}{\epsilon_0T_i}\right)$ where $n_e(n_i)$ is electron (ion) density, $e(q)$ is electron (ion) charge $T_e (T_i)$ is electron (ion) temperature. 
In experimental conditions, dust dynamics is also affected by the processes like dust-neutral collisions, ion-drag etc., however, the dominant interaction among dust particles remains the Yukawa potential. 

In dusty plasma, the sudden expansion of a closely packed clouds of charged dust particles has been observed in the plasma discharge experiments of Barkan and Merlino \cite{Barkan95}, Antonova {\it{et. al.}} \cite{antonova2012microparticles} and Merlino {\it{et. al.}} \cite{merlino2016coulomb}. Further, Ivlev et. al. \cite{ivlev2003decharging} experimentally studied the decharging of complex plasmas in the microgravity conditions and estimated the rest charge of dust particle at low gas pressure.
In these experiments, dust particles are initially confined into a densely packed cluster by the means of external confining potential. When the external confining potential is suddenly switching off, the cloud of dust particles tend to expand quickly or ``explode''. The later phase in which confining potential is off is known as the afterglow phase. In the afterglow phase the quantities like plasma density, plasma temperature and dust charge becomes function of time.
The MD simulation for the situation of afterglow plasma was performed by Saxena et. al. \cite{saxena2012dust} which demonstrates that the decay of plasma density and dust charge with time are critical function of background gas pressure. 
On switching off the power in low gas pressure regime, the plasma decays rapidly, therefore, the effective shielding of dust particles is also switched off. However, the charge on dust particle remains same as before switching off the power. Since, there is no shielding of dust potential (as there is no plasma in background), this leads to the Coulomb explosion of dust particles. 
In contrast to this, at higher background pressure ($>100$ mTorr), the plasma losses are slowed by neutrals and screening is preserved even in the afterglow phase which causes Yukawa explosion with screened dust charge. The hydrodynamic free expansion of charged particles (starting from an equilibrium state) has been examined analytically using fluid equations by Ivlev \cite{ivlev2013coulomb} and via MD simulations by Piel and Goree \cite{piel2013collisional} where they discussed the fundamental features of expansion for Coulomb balls and Yukawa balls separately in the collisional and collisionless limits. 

We, in this communication, describe the free, non quasi static expansion of Yukawa gas and subsequent change in thermodynamic variables in between different equilibrium states using first principle MD simulation. In our simulation, the dust charge and the Debye lengths are assumed to be constant. Hence, our simulation corresponds to a situation where background plasma conditions are not changing with time. In this respect, the present simulation is different from the expansion of dust particles in after glow phase described earlier. The MD results are also compared with the analytical results, however, instead of using hydrodynamic equations to study the evolution through non equilibrium states, we use thermodynamic equations to directly calculate the final equilibrium state from the initial equilibrium state.  

Paper is organized in the following manner. In Sec. \ref{Sec. II}, the details of MD simulation is described while simulation results are given in Sec. \ref{Sec. III}. The mean field solutions are derived in Sec. \ref{Sec. IV} where we also provide the validation of simulation results with mean field limit. The summary and possible application of our work in context of dusty plasma is discussed in Sec. \ref{Sec. V} where we also comment on the inversion temperature of Yukawa gas.

\section{Simulation Details \label{Sec. II}}
The equation of motion for $i^{th}$ particle is given by
\begin{equation}
 m_i \frac{d^2 r_i}{dt^2} = - \nabla_i \left( \sum_{j,~j\ne i}^{N_d} \phi(r_{ij})\right)
 \label{Eq:2}
\end{equation}
where $ r_{ij}= |\vec{r_i}-\vec{r_j}|$.
In our simulations, lengths are normalized by the background plasma Debye length $\lambda_D$, time is normalized by $\omega_0^{-1}$ where $\omega_0 = \left({Q_d^2}/{4\pi\epsilon_0 m  \lambda_D^3} \right)^{1/2}$ while the energies and temperature ($k_B T$) are measured in the units of $Q_d^2/4\pi\epsilon_0\lambda_D$, i.e.,
\begin{eqnarray}
  r &\rightarrow & \tilde{r}\lambda_D, \nonumber \\
  t & \rightarrow & \tilde{t} \omega_0^{-1}, \nonumber \\
  E & \rightarrow &  \tilde{E} \left(\frac{Q_d^2}{4\pi\epsilon_0\lambda_D }\right),\nonumber \\
  n_d & \rightarrow & \tilde n_d \lambda_D^3
  \label{Eq:3}
\end{eqnarray}
where $\tilde{r}$, $\tilde{t}$, $\tilde{E}$ and  $\tilde{n}_d$ represent the dimensionless length, time, energy and number density respectively. Thus, Eq. (\ref{Eq:2}) becomes,
\begin{equation}
\frac{d^2 \vec{\tilde r_i}}{d\tilde t^2}= \sum_{j,~j\neq i}^{N_d} (1+ \tilde r_{ij}) \frac{\exp(-\tilde r_{ij})}{ \tilde r_{ij}^3} \vec{ \tilde r}_{ij}
 \label{Eq:4} 
\end{equation}
Eq. (\ref{Eq:4}) is integrated using Leapfrog integrator scheme.

The equilibrium state of a Yukawa system is usually defined in terms of two dimensionless parameters \cite{hamaguchi1994thermodynamics}: $\kappa=a/\lambda_D $ ; the ratio of the mean inter-particle distance $ a=\left( 3/4\pi n_d\right)^{1/3} $ (here $n_d$ is the dust number density) to the screening length $\lambda_D$ and $\Gamma={Q_d^2}/{4\pi\epsilon_0 a T_d} $ ; the inverse of dust temperature $T_d$ measured in units of $ {Q_d^2}/{4\pi\epsilon_0 a}$. The coupling parameter $\Gamma^{*}= \Gamma \exp(-\kappa)$ which is the ratio of the mean inter-particle potential energy to the mean kinetic energy, is used as a measure of coupling strength in dusty plasmas. For $\Gamma^* \ll 1$ the correlation effects are almost negligible and Yukawa system behaves like an ideal gas. For $\Gamma^* \sim 1$, system behaves like an interacting Yukawa fluid and $\Gamma^* \gg 1$ corresponds to a condensed solid state where particles arrange themselves in a regular lattice form. The parameters $\Gamma$ and $\kappa$ can be related to normalized dust density and temperature as; 

\begin{equation}
 \kappa = \left(\frac{3}{4 \pi \tilde n_d}\right)^{1/3} ~~~~~~~ ~~~~\text{and} ~~~~\Gamma= \frac{1}{\kappa \tilde T_d}
  \label{Eq:5}
 \end{equation}

\subsection{Methodology \label{subsec.IIB}}
To realize the free expansion process in MD simulation we implement the following steps with perfectly reflecting boundary conditions.

\begin{figure*}[]
\begin{center}
\vspace{-0.20cm}
\includegraphics[scale=1.1]{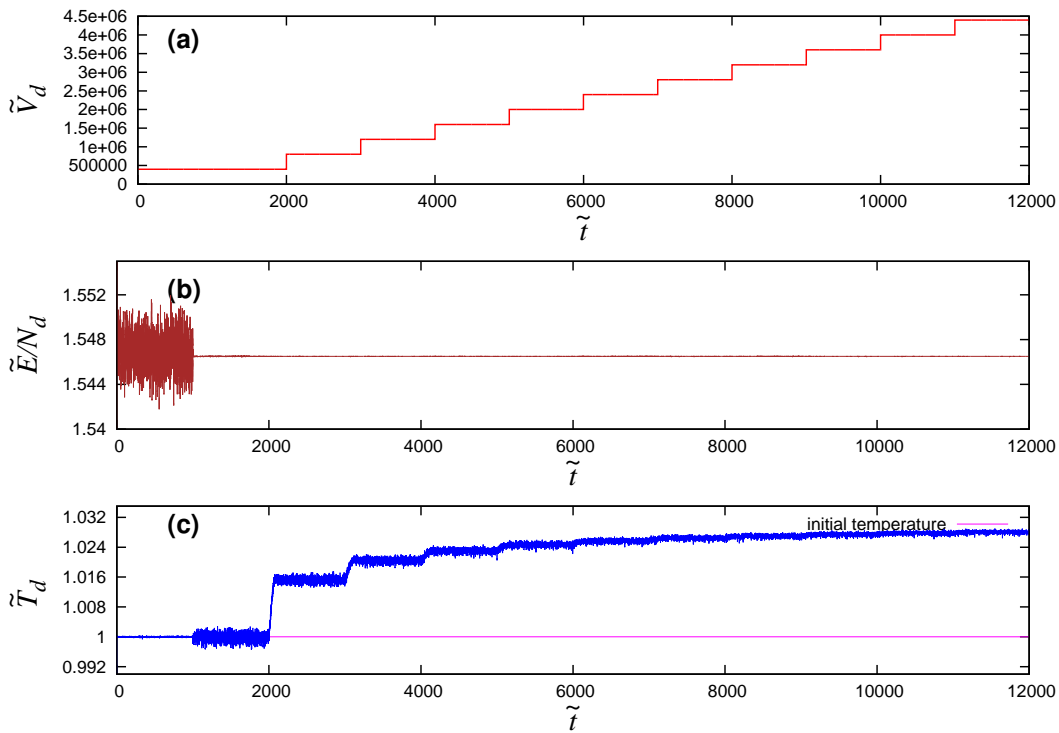}  
  \caption{\label{fig:1}
Time evolution of (a) volume, (b) total energy per particle, and (c) temperature is shown for the typical free expansion process of Yukawa gas for which initial temperature $\tilde T_{d0}=1.0$ and number density is $\tilde n_{d0}=0.01$.
}
 \end{center}
\end{figure*} 
\begin{enumerate}[(i)]
 \item We begin with an equilibrium state of known volume (say $V_i$), temperature ($T_i$) and fixed number of particles $N_d=4000$. The desired initial temperature is achieved by applying a Berendsen's thermostat for first $10^5$ time steps with each time step equals to $0.01~\omega_0^{-1}$. For the next  $10^5 $ time steps, system remains isolated where it reaches the equilibrium state.
 
 \item After step (i), the volume of system is increased suddenly by shifting one of the wall of simulation box so that the total volume becomes $V_i+V_0$ where $V_0$ is the increase in the volume (see Figure \ref{fig:1}). In the simulations the shift in wall is attained by changing the boundary conditions at a certain instant of time.
 
 \item The system is left undisturbed for $5 \times 10^4$ time steps so that it attains equilibrium in the new volume. After ensuring that the system is in equilibrium, the measurements for temperature and pressure is made over another $5 \times 10^4$ time steps.
 
 \item The expansion process is studied over a wide range of volume by repeating steps (ii) and (iii). 
 \item The reported temperature or pressure value is obtained by time average of these quantities over $5 \times 10^4$ time steps. The evolution of volume, total energy (i.e. the sum of kinetic and potential energies) per particle and temperature is shown in Figure \ref{fig:1}.

\end{enumerate}

\section{Simulation Results \label{Sec. III}}
Simulations are carried out for different initial conditions with fixed number of particles. 
Since, we are interested specifically in change of temperature of Yukawa gas in the free expansion process, we plot temperature as a function of volume. The plots of $\tilde T_{d}$ as function of $\kappa ~(=a/\lambda_D)$ are also presented to provide a better sense of parameters used in MD simulation. The results for initial temperature $\tilde T_d = 0.1,~ 0.5 $ and 1.0 are presented. The initial number density (before expansion) is kept same for different temperature runs and it's value is $\tilde n_{d0} =0.01$.

After several attempts, we were able to find a common scaling law for temperature as a function of volume which fits perfectly for all the range of parameters explored in this paper. The temperature is found to depend on volume as,
\begin{equation}
 \tilde T_d =   A-  B~\frac{N_d}{\tilde V_d}
 \label{fit-1}
\end{equation}
where $A$ and $B$ are fitting parameters determined by least square fitting. As number density $n_d$ (= $N_d/\tilde V_d$) is directly proportional to the cube of mean inter-particle distance (i.e., $n_d \propto a^3$), we can also fit our results as a function of $\kappa$ as
\begin{equation}
 \tilde T_d =   A-  \frac{C}{\kappa^3} \label{fit-2}
\end{equation}
where $C$ is another fitting parameter. 

\begin{figure}[]
\begin{center}
\subfigure[][]{\includegraphics[scale=0.6]{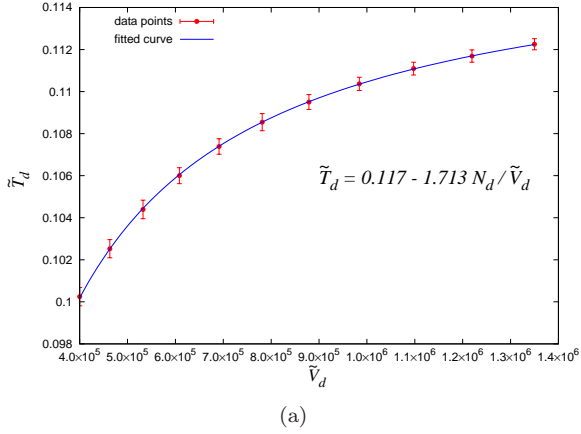}\label{fig:5.2a}}
\subfigure[][]{\includegraphics[scale=0.6]{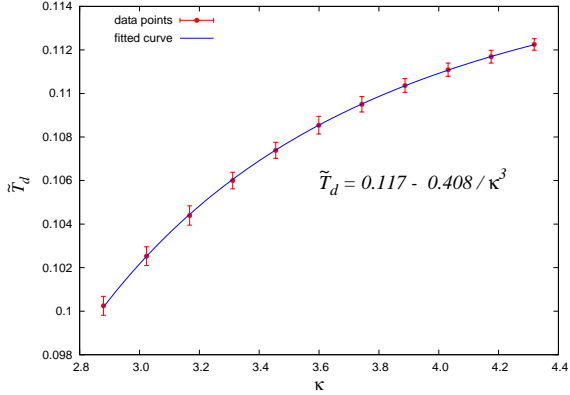}\label{fig:5.2b}} 
\caption{\label{fig:5.2} Temperature profile is shown as a function of normalized volume $\tilde V_d$ and $\kappa$ where initial temperature is $\tilde T_{d0}=0.1$ while initial number density is $\tilde n_{d0}=0.01$.   
}
 \end{center}
\end{figure} 
\subsection{Result for \boldmath{$\tilde T_{d0}=0.1$}}
MD results for initial temperature $\tilde T_{d0}=0.1$ are shown in Figure \ref{fig:5.2}. The values of fitting parameters of equation (\ref{fit-1}) \& (\ref{fit-2}) are  $A= 0.11731 \pm 2\times 10^{-5} $, $B = 1.71 \pm 0.004$ and $C = 0.40   \pm 0.001$. The temperature profile is given by
$$\tilde T_d =   0.11- 1.71 ~\tilde n_d ~~\text{or}~~ \tilde T_d=  0.11-  \frac{0.40}{\kappa^3}
$$
For the chosen set of parameters, the strong coupling parameter $\Gamma^*$ varies from 0.217 to 0.028. 

\begin{figure}[]
\begin{center}
\subfigure[][]{\includegraphics[scale=0.6]{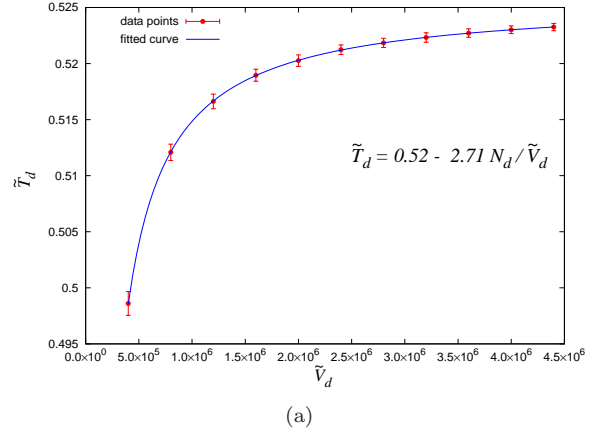}\label{fig:5.3a}}
\subfigure[][]{\includegraphics[scale=0.6]{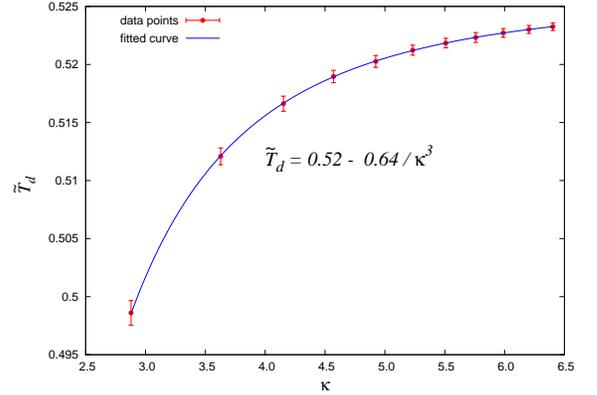}\label{fig:5.3b}}
  \caption{\label{fig:5.3} Temperature profile is shown as a function of normalized volume $\tilde V_d$ and $\kappa$ where initial temperature is $\tilde T_{d0}=0.5$ while initial number density is $\tilde n_{d0}=0.01$. 
}
 \end{center}
\end{figure} 

\subsection{Result for \boldmath{$\tilde T_{d0}=0.5$}}
The fitting parameters for initial temperature $\tilde T_{d0}=0.5$ (shown in Figure \ref{fig:5.3}) are $A= 0.52 \pm 1\times 10^{-6} $, $B = 2.71 \pm 0.005$ and $C = 0.64 \pm  0.001$. The functional dependence of temperature on $\tilde V_d$ and $\kappa$ is as follows;
$$\tilde T_d =   0.52- 2.71 ~\tilde n_d ~~\text{,}~~ \tilde T_d=  0.52 -  \frac{0.64}{\kappa^3}.
$$
The parameter $\Gamma^*$ varies from 0.033 to $4.4\times 10^{-4}$ in this run.

\begin{figure}[]
\begin{center}
\subfigure[][]{\includegraphics[scale=0.6]{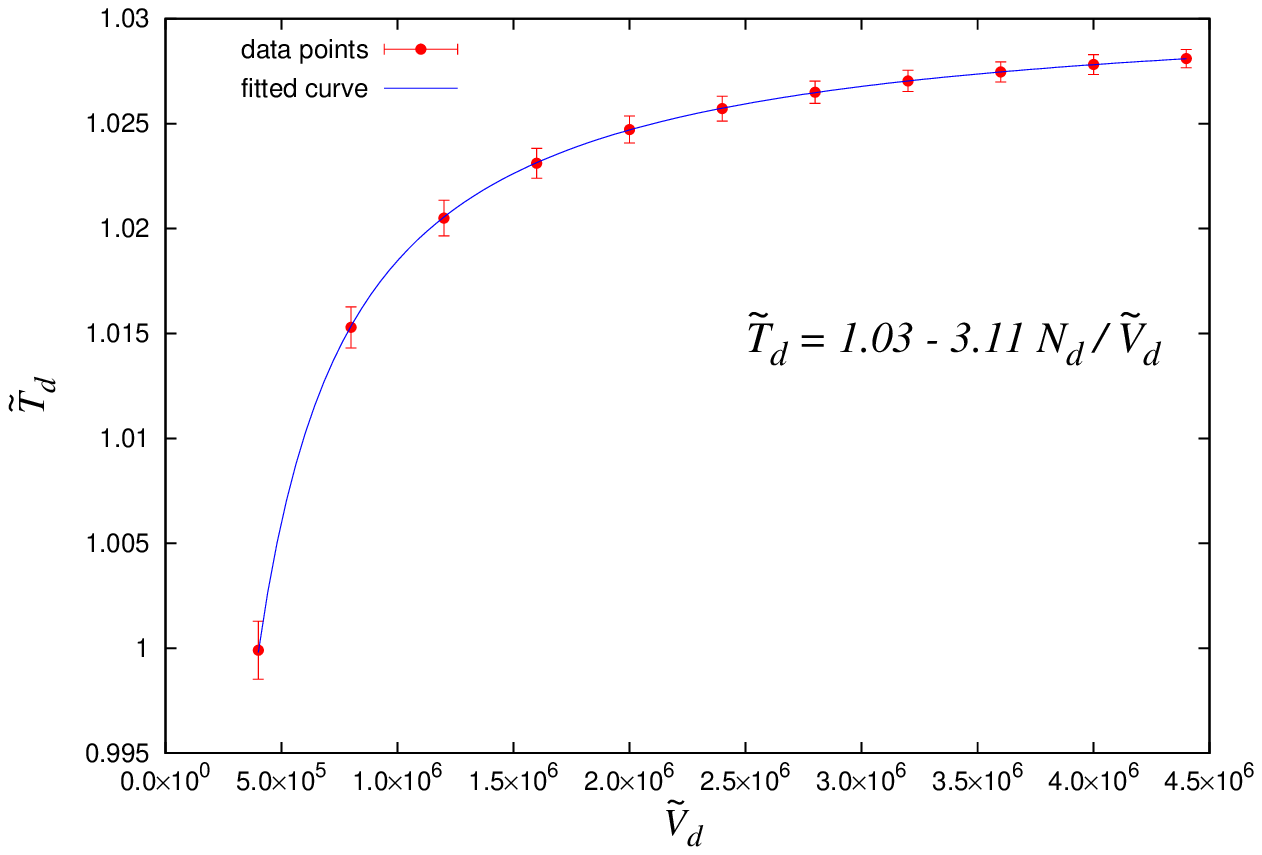}\label{fig:5.4a}}
\subfigure[][]{\includegraphics[scale=0.6]{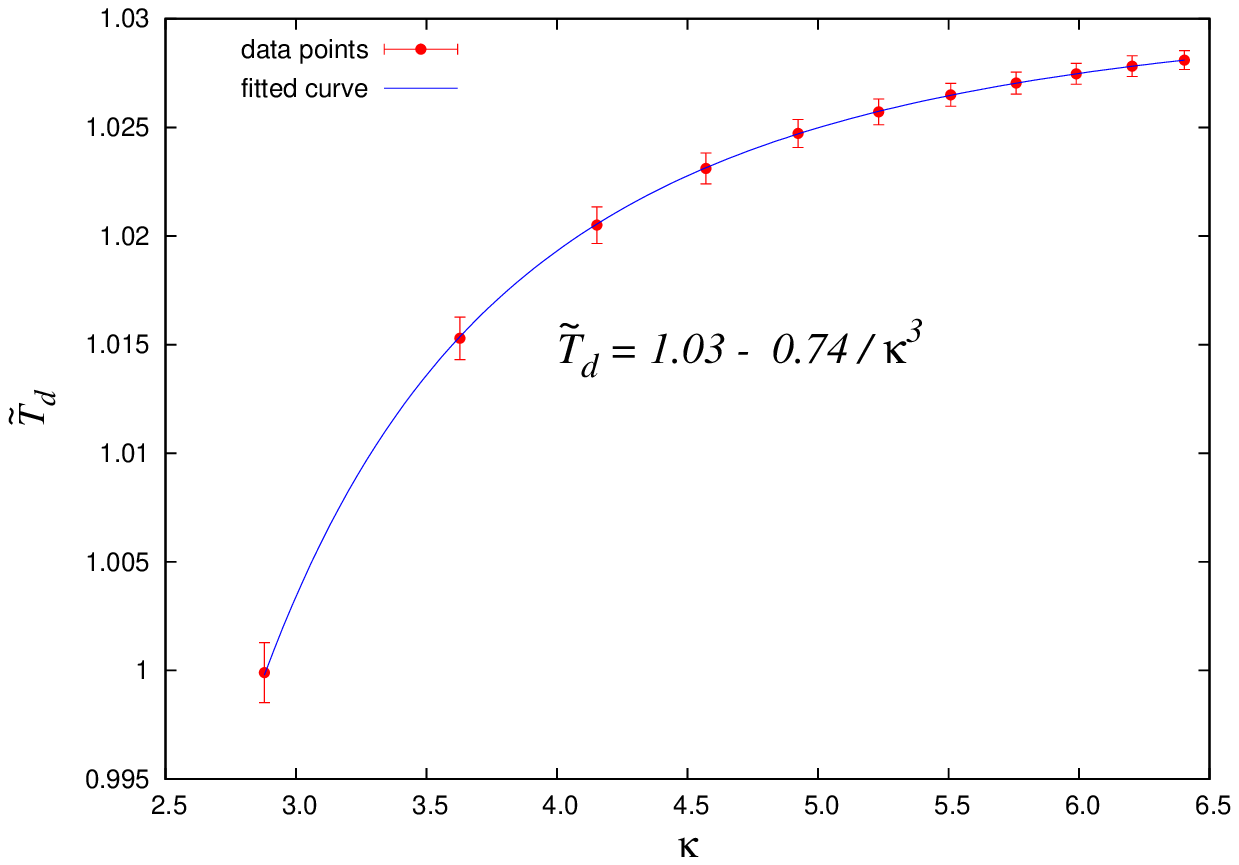}\label{fig:5.4b}}
  \caption{\label{fig:5.4} \small{Temperature profile is shown as a function of normalized volume $\tilde V_d$ and $\kappa$ where initial temperature is $\tilde T_{d0}=1.0$ while initial number density is $\tilde n_{d0}=0.01$.}  
}
 \end{center}
\end{figure} 
\subsection{Result for \boldmath{$\tilde T_{d0}=1.00$}}
The fitting parameters for initial temperature $\tilde T_{d0}=0.1$ (shown in Figure \ref{fig:5.4}) are $A= 1.03 \pm 1\times 10^{-5} $, $B = 3.11 \pm 0.006$ and $C = 0.74 \pm  0.001$. The functional dependence of temperature on $\tilde V_d$ and $\kappa$ is as follows;
$$\tilde T_d =   1.03- 3.11 ~\tilde n_d ~~\text{,}~~ \tilde T_d= 1.03 -  \frac{0.74}{\kappa^3}.
$$
The strong coupling parameter $\Gamma^*$ varies from 0.016 to $2.2\times 10^{-4}$.

\section{Validation of MD Results \label{Sec. IV}}
The observed scaling of temperature for the weakly coupled Yukawa gas can be explained simply on the basis of principle of energy conservation. 
The detailed thermodynamical model in context of dusty plasmas is discussed in the next section. For the system of particles interacting via Yukawa potential, the total energy $U$ is the sum of thermal energy and potential energy of particles.

\begin{equation}
 U= \frac{3}{2}N_d T_d + \frac{1}{2}\sum_{i=1}^{N_d}\sum_{j\neq i}^{N_d} \frac{Q_d^2}{4\pi\epsilon_0} \frac{\exp(-|r_i-r_j|/\lambda_D)}{|r_i-r_j|}.
 \label{Eq:5.7}
\end{equation}
In the weakly coupled homogeneous limit, the double summation of Eq. (\ref{Eq:5.7}) can be replaced with smooth integration\cite{avinash2010thermodynamics} over volume $V_d$ as $\sum_i^{N_d} \sum_j^{N_d} () = n_d N_d \int ()\;d\tau $ to give
\begin{eqnarray}
\frac{1}{2}\sum_{i=1}^{N_d}\sum_{j\neq i}^{N_d} \frac{Q_d^2}{4\pi\epsilon_0} \frac{\exp(-|r_i-r_j|/\lambda_D)}{|r_i-r_j|} 
&\rightarrow&  \frac{ Q_d^2\lambda_D^2 n_d N_d}{2\epsilon_0} X. 
\end{eqnarray}
where $X \left(=\int_{\bar r} \bar r\: \exp(-\bar r) d\bar{r} \right)$ is a dimensionless number to be fixed by MD simulation \cite{mks2017}. Substituting above results in equation(\ref{Eq:5.7}), we have expression for $U$ as,
\begin{equation}
 U= \frac{3}{2}N_d T_d +  \frac{ Q_d^2\lambda_D^2 n_d N_d}{2\epsilon_0} X.
\end{equation}
Since the internal energy remains constant during the free expansion process, therefore, $\Delta U =0 $ implies,
\begin{equation}
 \frac{3}{2}N_d \Delta T_d + X \frac{ Q_d^2\lambda_D^2  N_d}{2\epsilon_0} \Delta n_d =0.
\end{equation}
The above equation can be written in dimensionless units (using the relation given in Eq. (\ref{Eq:3})) to give
\begin{equation}
 \frac{3}{2} \Delta \tilde T_d + X \Delta \tilde n_d = 0.
\end{equation}
Writing the change in temperature and number density as, $\Delta \tilde T_d = (\tilde T_d- \tilde T_{d0}) $ and $\Delta \tilde n_d =(\tilde n_d- \tilde n_{d0})$, we have 
\begin{equation}
 \tilde T_d = \left(\tilde T_{d0} + \frac{2X}{3} \tilde n_{d0}  \right)- \frac{2X}{3} \tilde n_d.
 \label{Eq.5.10}
\end{equation}
The parameters $A~ \& ~B$ of used for fitting the MD results can be obtained from Eq. (\ref{Eq.5.10}) where $B=2X/3$  and $A$ is related to $B$ via initial temperature and number density
\begin{equation}
A = \tilde T_{d0} + B \tilde n_{d0}. \nonumber
\end{equation}
After knowing the relation between $\tilde T_d$ and $\tilde n_d$, the relation between  $\tilde T_d$ and $\kappa$ is straightforward.
Using the relation $4\pi a^3 n_d/3=1$ in dimensionless units, we get $\tilde n_d = 3/(4\pi \kappa^3)$. Thus, fitting parameter $C$ is related to $B$ as $C=3B/4\pi$. It should be noted that the relation among fitting parameters $A,B$ and $C$ can directly be verified from our simulations over a wide range of parameters presented in this paper.

\section{Discussion and Summary \label{Sec. V}}
In summary, we have presented a numerical model for free expansion of Yukawa gas using MD simulations and the results are explained on the basis of mean field theory. The free expansion process discussed in this paper could practically be realized in four component dusty plasmas experiments where interaction among dust grains is governed by Yukawa potential. It should be noted that in our model discussed so far we have modeled the ensemble of dust grains without taking into account the role of background plasma. However, in dusty plasma experiments the free expansion process involves not only the expansion of macroscopic dust particle but also the perturbation of gaseous electron ion plasma present in the background of dust grains. For typical dusty plasma experiments $n_e, n_i\sim 10^{14\-- 15} ~m^{-3}$, $n_d\sim 10^{9\-- 10} ~m^{-3}$ while neutral density $\sim 10^{19\--21} ~m^{-3}$, therefore, the heat capacity of neutrals gas is much more higher that of plasma and dust particles. Thus, neutral gas plays the role of heat bath for thermodynamic processes in dusty plasmas. At high neutral gas pressure ($\sim$ 100 $\--$ 200 mTorr) the dust$\--$neutral collisions are high, thus, dust maintains a good thermal contact with neutrals and dust component can be treated as isothermal. On the other hand, if the neutral gas pressure is low the thermal contact between dust and neutrals is weak and dust can be treated ``isolated'' for the thermodynamic processes. Contrary to the dust component, the thermal conductivity of background plasma is very high compared to dust, thus, plasma is always isothermal \cite{Avinash2017}.  

\begin{figure}[]
\begin{center}
\includegraphics[scale=0.5]{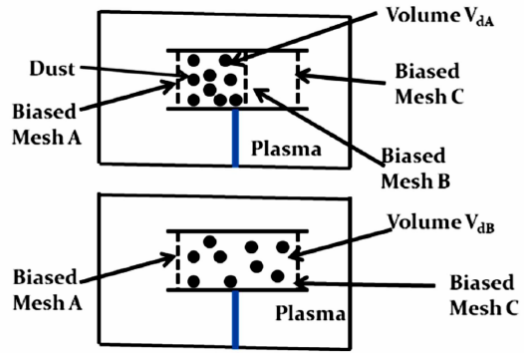}  
  \caption{\label{fig:5.5} Free expansion scheme for the dust in the plasma background is shown. The dust grains are initially confined in volume $V_{dA}$ between two biased meshes A and B as shown in figure. The whole arrangement is placed inside a cylindrical plasma discharge of volume V. At a later time, the biased of mesh B is removed and the dust particles are allowed to expand freely into bigger volume $V_{dB}$ between biased meshes A and C in the same plasma discharge. Figure is adapted from K. Avinash [private communications].}
 \end{center}
\end{figure}

The valid thermodynamic model which rightly explains the thermodynamic processes associated with dust grains in the constant plasma background is ``Particle within Plasma'' (PWP) model \cite{avinash2010thermodynamics,avinash2015theory}. In the PWP model, dust particles are confined in a finite volume $V_D$ within the plasma of much bigger volume $V (\text{where}~V \gg V_d)$ with the means of external fields and the two volumes, $V_d$ and $V$ could be varied independently. Different thermodynamic parameters of dust e.g., Helmholtz energy, pressure, entropy etc. using PWP model have been calculated in Ref.\onlinecite{avinash2010thermodynamics,avinash2015theory}. Since, plasma volume $V$ remains constant during free expansion of dust, the energy conservation for the quasi static work by $P_{ext}$ on dust volume $V_d$ demands that 
\begin{equation}
\Delta Q+\Delta Q_d=\Delta U+P_{ext}\Delta V_d,
\label{Eq:F1}
\end{equation}
where $\Delta Q ~ (= \Delta Q_e+\Delta Q_i)$ is the heat exchanged by plasma, $\Delta Q_d$ is heat exchanged by dust component and $U$ is the internal energy of the composite system. The internal energy $U$ for the composite system given by equation \cite{avinash2015theory},
\begin{eqnarray}
U = && \frac{3}{2}\sum_{\alpha=e,i,d} N_{\alpha}T_{\alpha} - \frac{3 Q_d^2 N_d}{16 \pi \epsilon_0\lambda_D} \nonumber \\
&+&  \frac{Q_d^2}{8 \pi \epsilon_0}\sum_{i=1}^{N_d}\sum_{j\neq i}^{N_d} \frac{\exp \left(-|r_i-r_j|/\lambda_D\right)}{|r_i-r_j|} \nonumber \\	
&-& \frac{Q_d^2}{8 \pi \epsilon_0}\sum_{i=1}^{N_d}\sum_{j\neq i}^{N_d}  \frac{\exp \left(-|r_i-r_j|/\lambda_D\right)}{2\lambda_D} . 
\label{Eq:F2}
\end{eqnarray}
For free expansion process work done against external pressure is zero i.e. $P_{ext}\Delta V_d=0$. Since, dust is thermally isolated $\Delta Q_d=0$. Now because thermal conductivity of plasma is very large, plasma temperature is constant i.e. $\Delta T_e=\Delta T_i=0$. Plasma absorbs a sufficient amount of heat from the heat bath in order to maintain the electron-ion temperature at constant values, which is given by \cite{Avinash2017}
\begin{equation}
 \Delta Q = \Delta\left(- \frac{Q_d^2}{8 \pi \epsilon_0}\sum_{i=1}^{N_d}\sum_{j\neq i}^{N_d} \frac{\exp \left(-|r_i-r_j|/\lambda_D\right)}{2\lambda_D} \right).
\end{equation}
With these substitutions, Eq. (\ref{Eq:F1}) becomes
\begin{equation}
 0=\frac{3}{2} N_d\Delta T_d + \Delta \left(\frac{Q_d^2}{8 \pi \epsilon_0}\sum_{i=1}^{N_d}\sum_{j\neq i}^{N_d} \frac{\exp \left(-k_D|r_i-r_j|\right)}{|r_i-r_j|}\right).
 \label{Eq:F3}
\end{equation}
This equation describes the change in temperature for dust component in the free expansion of dusty plasmas. We have evaluated this change using MD simulation in the present paper. The scheme of free expansion of dust particles in the background of gaseous plasma is shown in Figure \ref{fig:5.5}.

\begin{figure*}[]
\begin{center}
\mbox{\subfigure[][]{\includegraphics[scale=0.7]{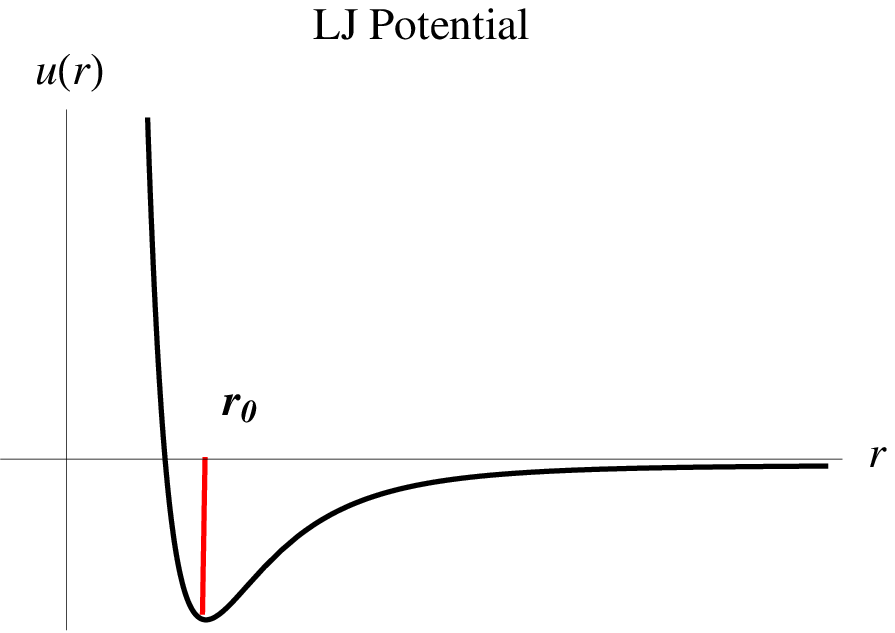}\label{fig:5.6a}}
\subfigure[][]{\includegraphics[scale=0.7]{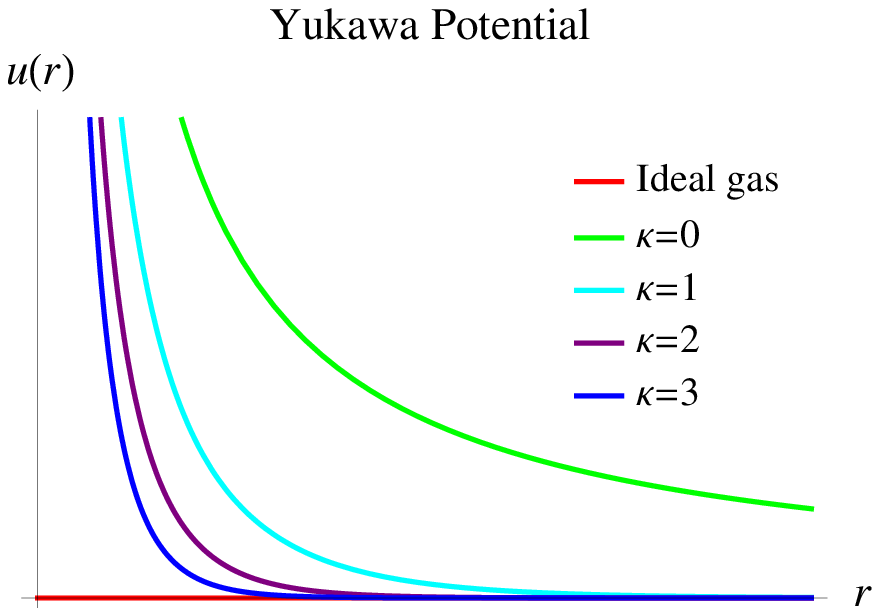}\label{fig:5.6b}}}  
 \caption{ The potential energy for two molecules as a function of $r$ is shown for real gas interacting via  (a) Lennard-Jones (LJ) potential and (b) Yukawa potential ($\exp(-\kappa r)/r$) for different values of $\kappa$. }
 \label{fig:5.6}
 \end{center}
\end{figure*} 
Another point we wish to emphasize here is that unlike the Van der Waals gases, Yukawa gas has no inversion temperature. Most of the real gases show cooling effect at room temperature and the sign of Joule's constant $J=(\partial T/\partial V)_U$ is negative. However, above a certain temperature $T_I$ called ``inversion temperature'' the real gases show heating effect and sign of $J$ becomes positive. The behavior of Van der Waal gases is described by Lennard-Jones (LJ) potential shown in Figure \ref{fig:5.6a}. The LJ potential is highly repulsive for short range intermolecular distances $r< r_0$ while it is attractive for distances $r>r_0$. At temperature $T<T_I$, the average intermolecular distance $\bar r$ of gas molecules is greater than $r_0$ so that overall interaction is attractive. Therefore, on expansion the potential energy increases while kinetic energy or temperature decreases as total energy remains constant. At higher temperature i.e., $T>T_I$ some of the molecules, due to their kinetic energy, come very close to each other so that intermolecular distance becomes smaller than $r_0$ while the average intermolecular distance may still be greater than $r_0$. The potential for $r<r_0$ is so high that overall interaction becomes repulsive and gas shows heating on free expansion\cite{Free-expansion93}.
For Yukawa potential, there is no such distance $r_0$ where interaction could change its sign (Figure \ref{fig:5.6b}) and Yukawa potential remains repulsive over all the ranges. This is why Yukawa gas shows heating effect irrespective of its temperature. 

\acknowledgments
MKS acknowledges  the financial support from University Grants Commission (UGC), India, under JRF/SRF scheme. MKS is thankful to Prof. R. Ganesh, IPR for his kind help regarding the MD simulation and providing the access to IPR's high performance cluster machines Uday and Udbhav  on which the major part of computing of this paper has been done.

%

\end{document}